# DeepSeek Powered Solid Dosage Formulation Design and Development

[1,2]Leqi Lin, [1,2]Xingyu Zhou, [1,2]Kaiyuan Yang, [1,2]*Xizhong Chen
[1]Department of Chemical Engineering, School of Chemistry and Chemical Engineering, Shanghai Jiao Tong University, Shanghai, China.
[2]State Key Laboratory of Synergistic Chem-Bio Synthesis, School of Chemistry and Chemical Engineering, Shanghai Jiao Tong University, Shanghai 200240, People's Republic of China

*Corresponding author Email: chenxizh@sjtu.edu.cn
Phone: +86-21- 54742893. Fax: +86-21-54745602.

**Abstract**
Pharmaceutical process design and development for generic, innovative, or personalized drugs have always been a time-consuming, costly, rigorous process, that involves multi-stage evaluation for better quality control and assurance. Large language models (LLMs), a type of generative artificial intelligence system, can augment laboratory research in the pharmaceutical engineering process by helping scientists to extract knowledge from literature, design parameters, and collect and interpret experimental data — ultimately accelerating scientific discovery. LLMs with prompt engineering technologies change the researcher's thinking protocol from traditional "empirical knowledge" to streamlined thinking that connects the performance and structured parameters together. In this work, we investigate and evaluate how prompt engineering technologies can enhance the drug design process from different strategies such as zero-shot, few-shot, chain-of-thought, etc. The dissolution profile for specific drugs is predicted and suggested from the LLMs model. Furthermore, the fundamental physical properties such as PSD, aspect ratio, and specific surface area could be inversely designed from the LLMs model. Finally, all the results are evaluated and validated by real-world cases to prove the reliability of prompt engineering techniques. Overall, there was a good agreement between the experimental dissolution data and the predicted dissolution profiles using the proposed model across hydrochlorothiazide powders. Initial evaluations show an MSE of 23.61 and $R^2$ of 0.97 in zero-shot, an MSE of 114.89 and $R^2$ of 0.90 in zero-shot-CoT, an MSE of 57.0 and $R^2$ of 0.92 in few-shot, a MSE of 22.56 and $R^2$ of 0.97 in few-shot-CoT and a MSE of 10.56 and $R^2$ of 0.99 with the involvement of RAG. This work breaks down any barriers in developing a systematic framework where LLMs assist in formulation design, process control, and decision-making. Finally, we conclude the work by discussing open challenges and future research directions in pharmaceutical processes.

**Highlights:**
1. an LLM-driven framework that Leverages prompt engineering to enhance solid

dosage design, and development and inversely optimizing key physical properties.
2. A fast and smart adaptive weighted retrieval equation to precisely select the data from RAG.
3. High efficiency on developing the dissolution profile and inverse designing the properties in comparison with experimental dissolution data and the predicted dissolution profiles.

**Introduction**

Pharmaceutical process engineering, as the backbone of drug product commercialization, contributes significantly to the early stages of drug development and formulation. Despite its pivotal role in ensuring drug efficacy and manufacturability, the field faces systemic challenges rooted in historical practices. Traditional development workflows—spanning formulation design, process parameter optimization, and quality control—often rely on empirical knowledge and iterative experimentation. For instance, critical parameters such as particle size distribution (PSD) thresholds or mixing time determination are frequently guided by legacy datasets rather than mechanistic models. This empirical dependency introduces significant variability; studies indicate that 25-40% of late-stage clinical failures stem from inconsistent dissolution profiles, necessitating costly mid-development recalibrations[1,2]. Moreover, it has been reported that "tremendous amount of attrition in the drug development process" (10- to 15-year discovery and development cycle that results in one new FDA-approved drug costs more than $2 billion)[3,4]. Thus, the pharmaceutical company urgently demands paradigm-shifting tools to systematize fragmented knowledge, decrease cost, and increase efficiency through formulation and clinical trials.

To address these challenges, conventional approaches have embraced computational simulations and machine learning (ML). Physics-based tools like discrete element modeling (DEM)[5-7] for powder blending or finite element analysis (FEA)[8,9] for the dissolution behavior of tablets provide mechanistic insights but struggle with multi-parametric systems and ideal assumptions. Besides, in the theoretically method, the results are usually varied (under or over-estimate the dissolution profile) due to the assumption of steady and un-steady state or translational diffusion and radiation diffusion, etc, which generate high deviation from the users' preferences[10-13]. Data-driven ML models, recently have been widely adapted to effective correlates input variables (e.g., API solubility, mixing time, particle size distribution) with outputs (e.g., flowability, dissolution, and hardness). Hornick et al. report generative AI as a tool for pharmaceutical particle engineering, offering a faster, cheaper, and more sustainable path to optimize drug formulations while maintaining rigorous performance standards[14]. Nevertheless, Data-driven ML models remain constrained by their hunger for labeled training data and costly pre-trained process[7,15,16]. These gaps highlight an unmet need to unify unstructured literature insights and experimental data into actionable optimization strategies.

Emerging advancements in large language models (LLMs) offer a groundbreaking avenue to bridge this divide. Modern LLMs like GPT-4 and DeepSeek, when coupled with prompt engineering technologies such as retrieval-augmented generation (RAG) and reinforcement learning from human feedback (RLHF), exhibit unprecedented capabilities in contextualizing domain-specific knowledge[17-20]. In materials science, such systems have demonstrated proficiency in tasks ranging from polymer property prediction to metal-organic framework (MOF) synthesis protocols[21-23]. Crucially, structured prompting techniques—such as chain-of-thought decomposition—enable LLMs to emulate expert reasoning, breaking down complex queries into sequential sub-tasks: distribution of particle size (geometry of particles) to environmental parameters and dissolution profiles. By grounding these outputs in scientific corpora (e.g., USP guidelines, patent databases, or literature), LLMs transcend conventional ML's data limitations, offering hypothesis-driven solutions even for understudied systems. Yet, their application to pharmaceutical process control—particularly in linking multivariate inputs (particle geometry, solubility, sink conditions) to kinetic outputs (dissolution profiles)—remains nascent, constrained by a lack of tailored prompting strategies.

This work pioneers a DeepSeek-LLMs framework specifically engineered for pharmaceutical process optimization. As illustrated in Fig. 1, our workflow synergizes three innovations: (1) Structured prompt templates that encode regulatory constraints and physicochemical relationships (e.g., Nernst-Brunner translation dissolution and radial diffusion dynamics)[10,24], enabling zero-shot inference of optimal parameters (e.g., predicting the impact of API aspect ratio on bioavailability); (2) data augmentation via RAG, where real-time retrieval from experimental and simulated dissolution databases; and (3) an adaptive weighted retrieval to iteratively refining prompts from RAG to minimize hallucinations. Initial evaluations show an MSE of 23.61 and $R^2$ of 0.97 in zero-shot, an MSE of 114.89 and $R^2$ of 0.90 in zero-shot-CoT, an MSE of 57.0 and $R^2$ of 0.92 in few-shot, a MSE of 22.56 and $R^2$ of 0.97 in few-shot-CoT and a MSE of 10.56 and $R^2$ of 0.99 with the involvement of RAG. Ultimately, this framework aims to dismantle silos between empirical expertise and data-driven innovation, ushering in an era of accelerated, first-pass-right development for next-generation personalized medication.

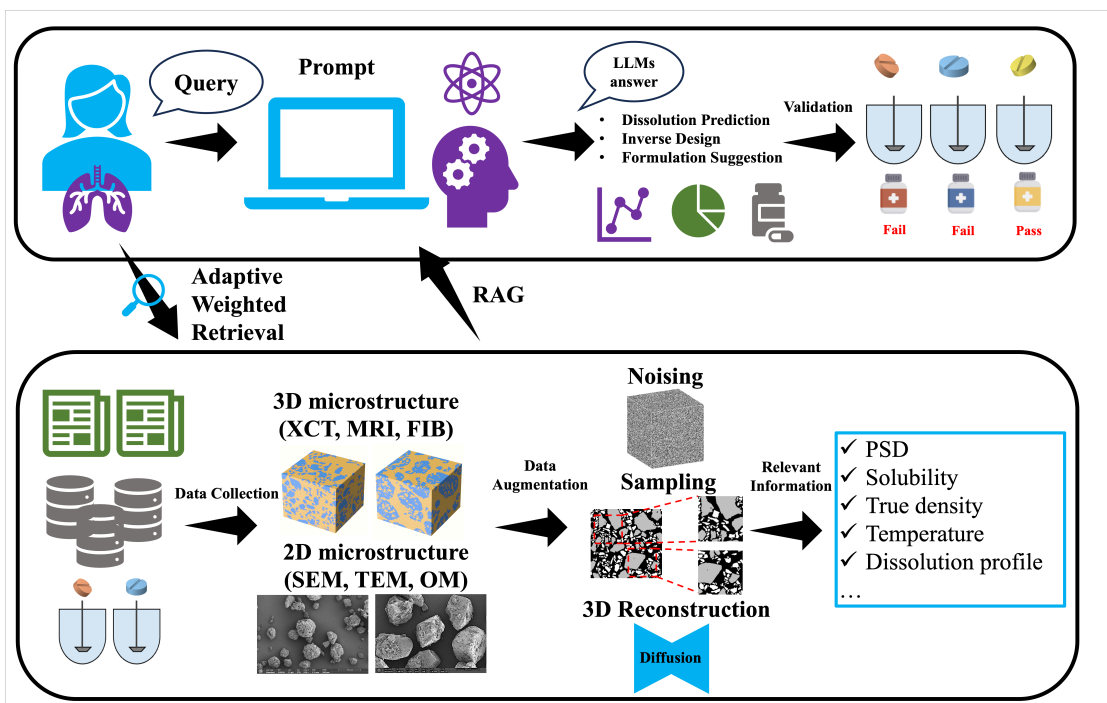

**Fig. 1.** Pipeline for LLMs driven Multimodal learning in solid dosage design and development.

## Methods

**Large Language Models**

The recent advances in transformer-based large language models (LLMs), pre-trained on Web-scale text corpora, significantly extended the capabilities of language models (LLMs) [25]. For instance, DeepSeek (V3, R1, or other distilled small models), fully open-source and commercialized freely AI models can be used not only for natural language processing but also as general task solvers that follow human instructions of complex new tasks performing multi-step reasoning when needed.

In this work, DeepSeek-R1-671b, a large-scale language model developed by DeepSeek Inc., featuring 671 billion parameters and designed for high-performance, multimodal artificial intelligence applications is utilized. DeepSeek-R1-671b demonstrates superior capabilities in multi-step problem-solving, cross-lingual processing, and real-time adaptation to user intent, making it a powerful tool for enterprises and researchers requiring precision and scalability.

**Evaluation Metrics**

Evaluations of the output results from LLMs are validated by experimental and simulation results obtained from our lab and from the literature (Salish et al. and Djuka et al. [10,24]). The release (dissolution) rate to time curves from the LLMs and literature are evaluated by using Mean Squared Error (MSE) and R-squared ($R^2$):

$$\text{MSE} = \frac{1}{n}\sum_{i=1}^{n}(y_i - \hat{y}_i)^2 \qquad (1)$$

where $y_i$ is the true value and $\hat{y}_i$ is the predicted value.

$$R^2 = 1 - \frac{\sum_{i=1}^{n}(y_i - \hat{y}_i)^2}{\sum_{i=1}^{n}(y_i - \bar{y})^2} \qquad (2)$$

where $\bar{y}$ is the mean of the observed data.

**Dissolution Experiments**

The dissolution tests are run followed by USP II paddle method with a dissolution bath described by USP General Chapter Dissolution <711>[26-28]. The assembly consists of the following: a vessel, a motor, a paddle formed from a blade and a shaft is used as the stirring element which may be covered, made of glass or other inert, transparent material. The shaft is positioned so that its axis is not more than 2 mm from the vertical axis of the vessel at any point and rotates smoothly without significant wobble that could affect the results. The vertical center line of the blade passes through the axis of the shaft so that the bottom of the blade is flush with the bottom of the shaft. The dosage unit is allowed to sink to the bottom of the vessel before rotation of the blade is started. A small, loose piece of nonreactive material, such as not more than a few turns of wire helix, may be attached to dosage units that would otherwise float.

Placing the stated volume of the dissolution medium (±1%) in the vessel in which the volume of medium and pH value is provided by the FDA Dissolution Methods Database to meet the standard requirements of United States of America Pharmacopoeia. Assembling the apparatus, equilibrate the dissolution medium to 37 ± 0.5°, and remove the thermometer. Place 1 dosage unit in the apparatus, taking care to exclude air bubbles from the surface of the dosage unit, and immediately operate the apparatus at the specified rate. Within the time interval specified, or at each of the times stated, withdraw a specimen from a zone midway between the surface of the dissolution medium and the top of the rotating basket or blade, not less than 1 cm from the vessel wall. Where multiple sampling times are specified, replace the aliquots withdrawn for analysis with equal volumes of fresh dissolution medium at 37° or, where it can be shown that replacement of the medium is not necessary, correct for the volume change in the calculation. Keep the vessel covered for the duration of the test, and verify the temperature of the mixture under test at suitable times. Repeat the test with additional dosage form units.

The solid dosage forms of hydrochlorothiazide powders were run in pH 7.2 phosphate buffer to meet the standard requirements of United States of America Pharmacopoeia. The concentration of drugs is analyzed by UV-vis spectrometer. For each sample, the dissolution testing was performed in five replicates.

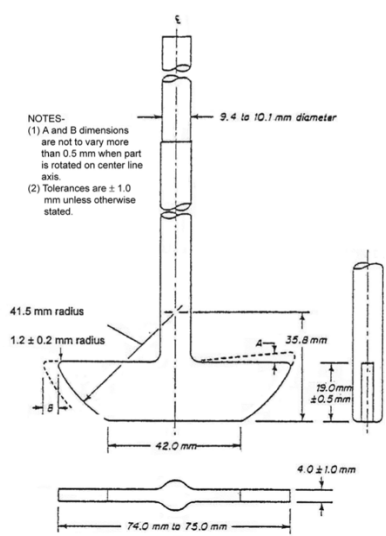

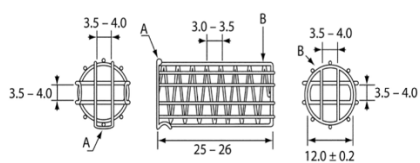

Fig. 2. The schematic of USP II paddle method for dissolution test[26-28].

## Results and Discussion
### Forms Design and Development from Prompt Engineering Techniques

Fig. 3 shows the structured prompt for zero-shot and zero-shot chain-of-thought (CoT) respectively, while CoT is to ask the LLMs to "Do the step-by-step analysis ". The examples of few-shot prompting as shown in Fig. 3 are real-experimental data extracted from literature provided by Salish et al[10]. In few-shot prompting, LLMs are instructed to perform specific experimental data analyses or interpretations, ensuring the model replicated methodologies or formats used in their research. By grounding the model in real-world examples, the authors could steer it toward precise, context-aware outputs aligned with their study's objectives. By integrating CoT, LLMs are now a collaborative reasoner that aligns with the proposed analytical workflow. For example, CoT breaks problems into smaller sub-missions that reduce the likelihood of mistakes, as the model validates each intermediate result. CoT can be done by explicitly prompting the model to think step by step (zero-shot-CoT) or include 1–3 examples in the prompt that illustrate the reasoning process (few-shot-CoT). This helps iteratively refine the model and review the early mistakes when happen.

### Role: ###
Hi, You are an expert on the Drug development.
You can design the particle size distribution for customed dissolution profiles,
or predict the Drug Released (%) based on given physical proerties such as particle size distribution.

### Background: ###
You have a bunch of experience on that and
have studied those commonly used emperical diffusion models
such as Nernst-Brunner translation dissolution and radial diffusion dynamics from
(1) Salish, K., So, C., Jeong, S. H., Hou, H. H. & Mao, C. A Refined Thin-Film Model for Drug Dissolution Considering Radial Diffusion - Simulating Powder Dissolution. Pharm Res 41, 947-958 (2024).
https://doi.org:10.1007/s11095-024-03696-0
(2) Djukaj, S., Kolar, J., Lehocky, R., Zadrazil, A. & Stepanek, F. Design of particle size distribution for custom dissolution profiles by solving the inverse problem. Powder Technology: An International Journal on the Science and Technology of Wet and Dry Particulate Systems, 395 (2022).

### Reqeust: ###
1. Your customer will give you several fundamental parameters and based on the given parameters,
2. you need to either predict the Drug Released (%) for the customer or
3. you need to design the physical properties of the drugs and optimize the conditions based on given dissolution profile (dissolution rate)

### Input Format: ###
{
  Input = {
    "Mean Particle Size, D50" : 97.5
    "Aspect ratio" : 1.0,
    "Roundness": 1.0,
    "solubility of  drug (mg/mL) " : 0.45,
    "Diffusion coefficient of drug (m^2/s)" :  7.5x 10^(-10),
    "True Density of drug   (g/mL)   " : 1.512,
    "Specific surface area (m^2/g)" : 1.07,
    "volume-based equivalent particle size (micrometer)" : 1.85,

  }
}

### Outout Format: ###
please generate a table with columns: [Time(min), Drug Released (%)].
Include key metrics: $t_{0}$, $t_{0.25}$, $t_{0.5}$, $t_{0.75}$, $t_{1}$, $t_{2}$, $t_{3}$, $t_{4}$, $t_{5}$, $t_{6}$
where t refers to the abbreviation of "Time (hrs)"

{
    "columns": ["Time (hr)", "Drug Released (%)"],
    "data": [
        [0, 0],
        [0.25, 85],
        [0.5, 87],
        [0.75, 88],
        [1, 89],
        [2, 89],
        [3, 89],
        [4, 88],
        [5, 87],
        [6, 87]
    ]
  }
}

### Examples: ###
no examples provided

### Constrains: ###
1. Nernst-Brunner equation = {
$$\frac{dx}{dt} = -\frac{k \psi_A}{3 \rho_s \psi_v} (C_{\text{sat}} - C_b)$$
Where $k = \frac{\text{Sh} \cdot D}{x}$,
$\text{Sh} = 2 + 0.52 \text{Re}^{0.52} \text{Sc}^{1/3}$
}
2. Final dissolution ≥85% within 60 min (USP compliance).
3. Please Do not make up recommendations without scientific basis.
4. Only provide optimizations that have clear scientific reasoning.
5. Do not make up the answer randomly if you may not be able to provide the correct answer.

**Fig. 3.** JSON structured prompt (zero-shot)

In this study, DeepSeek 671b is employed for drug dissolution profile design using structured prompt engineering to predict drug release versus time curves under specified conditions. Fig. 4 shows the LLMs results for zero-shot, zero-shot chain-of-thought (CoT), few-shot, and few-shot CoT respectively. Table 1 lists the MSE and $R^2$ evaluation on the LLMs' generated results and real data.

Zero-shot (ZS) prompt approach demonstrated a strong baseline performance with low error (MSE = 23.61%) and high explanatory power ($R^2$ = 0.97), indicating its effectiveness for tasks requiring minimal guidance. However, integrating Chain-of-Thought (CoT) into ZS (ZS_CoT) degraded performance significantly (MSE = 114.89%, $R^2$ = 0.90), which is likely due to ungrounded reasoning steps or hallucinations in the absence of contextual examples. Similarly, the few-shot (FS) prompt underperformed compared to ZS (MSE = 57.0%, $R^2$ = 0.92), suggesting that limited examples may introduce noise or misalignment with the task.

Notably, combining few-shot prompts with CoT (FS_CoT) dramatically improved accuracy, achieving near-ZS-level $R^2$ (0.97) while reducing MSE by 60% (22.56%). This highlights the critical role of task-specific examples in anchoring the CoT process to valid reasoning steps. The best performance, however, was achieved through Retrieval-Augmented Generation (RAG), which leverages external data from scientific literature or databases. RAG produced the lowest error (MSE = 10.55%) and highest explanatory power ($R^2$ = 0.99), underscoring its ability to ground predictions in real-world evidence and minimize speculative errors.

These findings emphasize that structured guidance—whether through examples, step-by-step reasoning, or external knowledge integration—is essential for optimizing LLM performance in complex scientific tasks like drug dissolution modeling. For practical applications, RAG emerges as the gold standard for precision, while FS_CoT offers a viable alternative when external data access is constrained.

### Few-shot Examples ###
### Example1: ###
### Input : ###
```
{
    Input = {
        "Mean Particle Size, D50" : 45
        "Aspect ratio" : 1.0,
        "Roundness": 1.0,
        "solubility of  drug (mg/mL) " : 0.45,
        "Diffusion coefficient of drug (m^2/s)" :  7.5x 10^(-10),
        "True Density of drug   (g/mL)  " : 1.512,
        "Specific surface area (m^2/g)" : 1.70,
        "volume-based equivalent particle size (micrometer)" : 1.17,
    }
}
```
### Outout: ###
```
{
    "columns": ["Time (hr)", "Drug Released (%)"],
    "data": [
        [0, 0],
        [0.25, 85],
        [0.5, 87],
        [0.75, 88],
        [1, 89],
        [2, 89],
        [3, 89],
        [4, 88],
        [5, 87],
        [6, 87]
    ]
}
```
### Example2: ###
### Input : ###
```
{
    Input = {
        "Mean Particle Size, D50" : 200,
        "Aspect ratio" : 1.0,
        "Roundness": 1.0,
        "solubility of  drug (mg/mL) " : 0.45,
        "Diffusion coefficient of drug (m^2/s)" :  7.5x 10^(-10),
        "True Density of drug   (g/mL)  " : 1.512,
        "Specific surface area (m^2/g)" : 0.24,
        "volume-based equivalent particle size (micrometer)" : 8.14,
    }
}
```
### Outout : ###
```
{
    "columns": ["Time (hr)", "Drug Released (%)"],
    "data": [
        [0, 0],
        [0.25, 12],
        [0.5, 20],
        [0.75, 28],
        [1, 35],
        [2, 52],
        [3, 63],
        [4, 71],
        [5, 75],
        [6, 82]
    ]
}
```
### Example3: ###
### Input : ###
```
{
    Input = {
        "Mean Particle Size, D50" : 97.5
        "Aspect ratio" : 1.0,
        "Roundness": 1.0,
        "solubility of  drug (mg/mL) " : 0.45,
        "Diffusion coefficient of drug (m^2/s)" :  7.5x 10^(-10),
        "True Density of drug   (g/mL)  " : 1.512,
        "Specific surface area (m^2/g)" : 0.16,
        "volume-based equivalent particle size (micrometer)" : 11.94,
    }
}
```
### Outout : ###
```
{
    "columns": ["Time (hr)", "Drug Released (%)"],
    "data": [
        [0, 0],
        [0.25, 32],
        [0.5, 40],
        [0.75, 56],
        [1, 60],
        [2, 72],
        [3, 76],
        [4, 80],
        [5, 82],
        [6, 87]
    ]
}
```

**Fig. 4.** Few-shot prompting examples. The input and output data are extracted from Salish et al.[10].

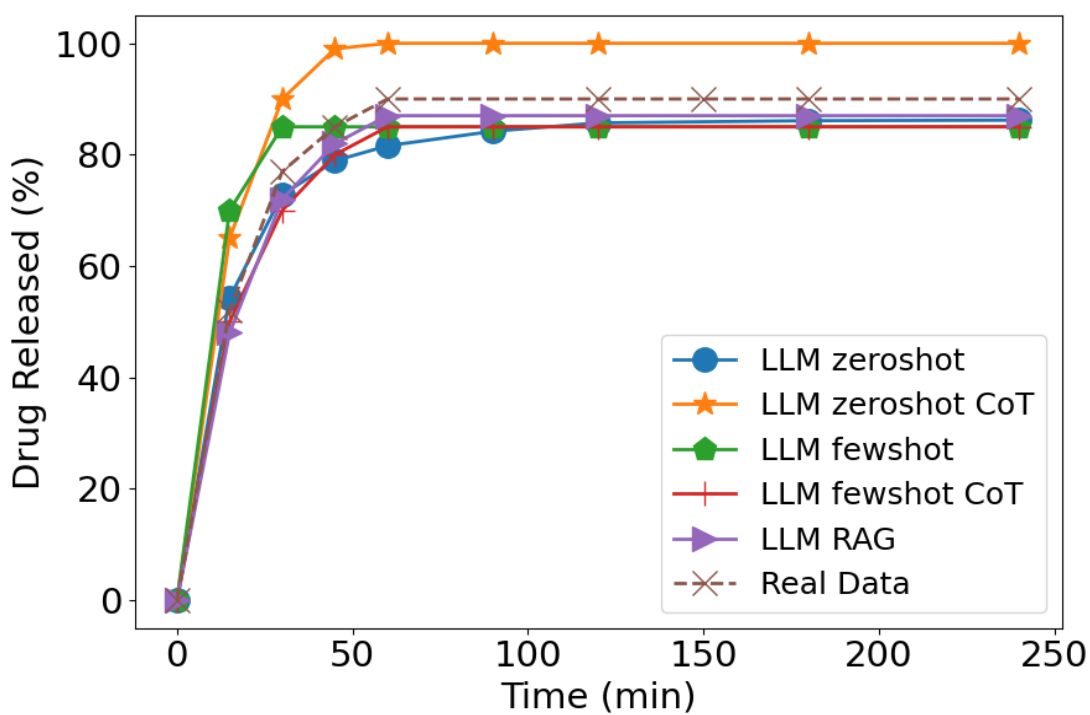

**Fig. 5.** LLMs results with the real-experimental data.

**Table 1.** MSE and R square comparison of LLMs results with the real-experimental data. (ZS: zero-shot, FS: Few-shot, CoT: Chain-of-Thought, RAG: retrieval-augmented generation)

|         | ZS    | ZS_CoT | FS   | FS_CoT | RAG   |
|---------|-------|--------|------|--------|-------|
| MSE (%) | 23.61 | 114.89 | 57.0 | 22.56  | 10.55 |
| R square| 0.97  | 0.90   | 0.92 | 0.97   | 0.99  |

**Challenges and Future Directions**

The integration of large language models (LLMs) with prompt engineering into pharmaceutical process design faces several challenges rooted in data limitations, interpretability, and systemic integration. A primary hurdle is the scarcity and inconsistency of domain-specific datasets, such as dissolution profiles or particle size distributions, which are often small, proprietary, or noisy, limiting the models' ability to generalize across diverse formulations. Additionally, the "black-box" nature of LLMs complicates trust and regulatory compliance, such as inverse-designed physical properties or dissolution behaviors that require transparent alignment with empirical pharmaceutical principles.

In the future directions, a self-constructed database with 30% real-experiment data and 70% simulated data will be built as RAG to enhance the LLMs' capability and reliability. The prompt engineering being developed now only includes the design of the dissolution profile and particle size distribution, however, flowability and mechanical properties are two of the most important factors affecting the product stability and quality. Thus, these two factors and other potential factors will be included and considered in the prompt engineering as well to broaden the capability of LLMs in drug design and development.

Ultimately, advancing LLMs from assistive tools to certified decision-making partners demands interdisciplinary innovation, uniting AI researchers, pharmaceutical engineers, and regulators to harmonize data-driven agility with the rigor of pharmaceutical science.